\documentstyle[prc,aps,epsfig]{revtex}

\newcommand{\etal}{\mbox{\textit et al.}}			%

\def\s{\mbox{\boldmath $s$}}
\def\h{\mbox{\boldmath $h$}}
\def\S{\mbox{\boldmath $S$}}
\def\N{\mbox{\boldmath $N$}}
\def\L{\mbox{\boldmath $L$}}

\def\p{\mbox{\boldmath $p$}}

\def\P{\mbox{\boldmath $P$}}
\def\q{\mbox{\boldmath $q$}}

\def\N{\mbox{\boldmath $N$}}

\def\ss{\mbox{\boldmath $\sigma$}}
\def\rrho {\mbox{\boldmath $\rho$}}

\begin{document}
\draft

\title{{POLARIZATION OBSERVABLES IN (${\vec e},e'{\vec N}N$) and 
(${\vec \gamma},{\vec N}N$) REACTIONS}\footnote[2]{submitted to the 5th Workshop 
on "e.-m. induced two-hadron emission", Lund, June 13-16, 2001.}}
\author{C.~Giusti and F.~D.~Pacati}
\address
 {Dipartimento di Fisica Nucleare e Teorica,
        Universit\`a di Pavia,\\ and 
	Istituto Nazionale di Fisica Nucleare, Sezione di Pavia, Pavia, 
	Italy}

\maketitle

\begin{abstract}
Nucleon recoil polarization in electromagnetic reactions with two-nucleon 
emission is discussed for both (${\vec e},e'{\vec N}N$) and 
(${\vec \gamma},{\vec N}N$). Numerical results are given for exclusive 
two-nucleon knockout reactions from $^{16}$O in a theoretical model where
final-state interactions, one-body and two-body currents, and the effect of 
correlations in the initial pair wave function are included.
\end{abstract}
\maketitle

\section{Introduction}

Electromagnetically induced two-nucleon knockout has long been devised as the 
most direct tool for exploring the properties of nucleon pairs within nuclei, 
their interaction at short distance and therefore short-range correlations 
(SRC)~\cite{Oxford}. In order to extract this information, a reliable 
theoretical model is needed, able to keep reasonably under control the 
reactions processes and the various ingredients, as well as a combined 
experimental study of different reactions in different situations, where the 
various contributions play a different role and can thus be disentangled and
separately investigated. 

Two nucleons can be naturally ejected by two-body currents, which effectively 
take into account the influence of subnuclear degrees of freedom like mesons 
and isobars. Direct insight into SRC can be obtained from the process where the 
real or virtual photon hits, through a one-body current, either nucleon of a 
correlated pair and both nucleons are then ejected from the nucleus. A reliable 
and consistent treatment of these two competing processes, both produced by the 
exchange of mesons between nucleons, is needed. Their role and relevance, 
however, can be different in different reactions and kinematics. It is thus 
possible to envisage situations where either process is dominant and can be 
separately investigated.

Interesting and complementary information is available from electron and
photon-induced reactions. The electron probe is however preferable for the study 
of SRC. In fact, two-body currents predominantly contribute to the transverse
components of the nuclear response. Only these components are present in
photon-induced reactions that appear thus generally dominated by two-body
currents. Also the longitudinal component, dominated by correlations, is present 
in electron-induced reactions. The possibility of independently varying 
the energy and momentum transfer of the exchanged virtual photon allows one to 
select kinematics where the longitudinal response and thus SRC are dominant. 

A combined study of $pp$ and $np$ knockout is needed for a complete information. 
Correlations are different in $pp$ and $np$ pairs. They are stronger in $np$ 
pairs and thus in $np$ knockout due to the tensor force, that is predominantly 
present in the wave function of a $np$ pair. But also two-body currents are much 
more important in $np$ knockout, while they are strongly suppressed in $pp$ 
knockout, where the charge-exchange terms of the two-body current do not 
contribute. Therefore, the ($e,e'pp$) reaction was devised as the preferential 
process for studying SRC in nuclei. It is however clear that, since different
effects can be emphasized in suitable conditions for different reactions, 
a combined study of $pp$ and $np$ knockout induced by real and virtual photons 
is needed to unravel the different contributions and obtain a clear and 
complete information on $NN$ correlations. 

Exclusive reactions, for transitions to specific discrete eigenstates of the
residual nucleus, are of particular interest for this study. One of the main
results of the theoretical investigation is the selectivity of exclusive 
reactions involving different final states that can be differently affected by 
one-body and two-body currents~\cite{giu98,pn}. Thus, the experimental 
resolution of specific final states may act as a filter to disentangle the two 
reaction processes. $^{16}$O is a suitable target for this study, due to the 
presence of discrete low-lying states in the experimental spectrum of $^{14}$C 
and $^{14}$N well separated in energy. From this point of  view, $^{16}$O is 
better than a light nucleus, which lacks specific final states. 

First measurements of the exclusive $^{16}$O($e,e'pp$)$^{14}$C reaction 
performed at NIKHEF~\cite{Gerco,NIKHEF,Ronald} and MAMI~\cite{Rosner,Rosner1} 
have basically confirmed the predictions of the theoretical model and have 
given clear evidence for SRC for the  transition to the ground state of 
$^{14}$C. This result gives rise to the hope that a similar separation between 
two-body currents and correlations is possible also in the ($e,e'np$) reaction. 
This would be of particular interest for the study of tensor correlations. No 
data are available at present, but a proposal for the experimental study of the 
exclusive $^{16}$O($e,e'np$)$^{14}$N reaction was approved in Mainz and is in 
preparation~\cite{MAMI}. 

First pioneering ($\gamma,NN$) experiments were performed in Bonn~\cite{Bonn} 
and Tokyo~\cite{Tokyo} with poor statistical accuracy and energy resolution. 
Further experiments carried out at MAMI achieved much better 
results~\cite{Mainz}. These experiments, with a resolution of about 6-9 MeV, 
were able to resolve the major shells from which the two nucleons are 
emitted. The first high-resolution $^{16}$O($\gamma,np$)$^{14}$N experiment able 
to separate final states was performed at MAXlab in Lund~\cite{Lund}, in a small 
energy range and with low statistics. A high-resolution experiment for the 
$^{16}$O($\gamma,np$)$^{14}$N and $^{16}$O($\gamma,pp$)$^{14}$C reactions in the 
photon-energy range between 90 and 270 MeV, aiming not only at separating final 
states but also at determining the momentum distributions of each state,  
was approved in Mainz and is in preparation~\cite{Mainz1}. Results for a first 
test and feasibility study are presented in~\cite{Grab}.

Good opportunities to increase the richness of information available from
two-nucleon knockout are offered by polarization measurements. Reactions with 
polarized particles give access to a larger number of observables, hidden in the 
unpolarized case and whose determination can impose more severe constraints on 
theoretical models. Some of these observables are expected to be sensitive to 
the small components of the transition amplitudes, which are generally masked by 
the dominant ones in the unpolarized cross section. These small components 
often contain interesting information on subtle effects and may thus represent 
a stringent test of theoretical models. This is the place where polarization 
observables enter, because in general they contain interference terms of the 
various matrix elements in different ways. Thus, a small amplitude may be 
considerably amplified by the interference with a dominant one. 

The asymmetry of the cross section for linearly polarized photons in 
($\gamma,pp$) and ($\gamma,pn$) was studied in~\cite{BGPR,Gent2}. 
Numerical results for exclusive $pp$ and $pn$ knockout from $^{16}$O can be 
found in~\cite{gnn}. First measurements with polarized photon beams have been 
performed at LEGS on $^3$He~\cite{LEGSHe} and $^{16}$O~\cite{LEGSO} and at MAMI 
on $^{12}$C~\cite{MacGre}. In these experiments, however, the energy resolution 
was not enough to separate specific final states of the residual nucleus. 

The case of the nucleon recoil polarization in (${\vec e},e'{\vec N}N$) and
(${\vec \gamma},{\vec N}N$) reactions where also the incident electron or the
incident photon is polarized is considered here. The formalism and the
polarization observables for (${\vec e},e'{\vec N}N$) and 
(${\vec \gamma},{\vec N}N$) are discussed in sects.~2 and 3, respectively. Some 
numerical examples for exclusive knockout reactions on $^{16}$O are presented 
in sect.~4. Other examples and a more detailed discussion of the formalism are 
given  in~\cite{pppol,gpnpol}. Some numerical results obtained with a different 
theoretical model can be found also in~\cite{Gent2,Gent1}. 

\section{Formalism for the 
(${\vec{\lowercase{e}}},{\lowercase{e}}'{\vec{N}}N$) reaction}

The triple coincidence cross section for the electron induced reaction where two
nucleons, with momenta $\p'_1$ and $\p'_2$ and energies $E'_1$ and $E'_2$, are 
emitted is obtained from the contraction between the lepton tensor
and the hadron tensor as a linear combination of 9 structure
functions~\cite{Oxford,GP}
\begin{eqnarray} 
\frac{{\mathrm d}^5\sigma} {{\mathrm d}E'_0 {\mathrm d}\Omega'_0  \,
{\mathrm d}\Omega'_1 \/{\mathrm d}\Omega'_2 \/{\mathrm d}E'_1} = & K & 
\Big \{ \epsilon_{\mathrm{L}} f_{00} + f_{11} + \sqrt{\epsilon_{\mathrm{L}}
(1+\epsilon)} (f_{01} \cos\alpha + \overline f_{01} \sin\alpha)  
- \epsilon  (f_{1-1} \cos2\alpha  \nonumber \\
& + & \overline f_{1-1} \sin2\alpha) 
  + h \Big [ {\sqrt{\epsilon_{\mathrm{L}} (1-\epsilon)}} 
(f'_{01} \sin\alpha + \overline f'_{01} \cos\alpha) + \sqrt{1-\epsilon^2} 
f'_{11} \Big ] \Big \}, \label{eq:cssec}
\end{eqnarray} 
where $E'_0$ is the energy of the outgoing electron and $K$ is a kinematic 
factor. The combination coefficients are obtained from the components of the
lepton tensor and depend only on electron kinematics, with 
\begin{equation}
\epsilon = \left( 1 + 2 \frac {\vert \q \vert^2} {Q^2} \tan^2 \frac {\theta}
{2} \right)^{-1} , \,\,  \epsilon_{\mathrm L} = 
\frac {Q^2} {\vert \q \vert^2} \epsilon,
\label{eq:eps}
\end{equation}
where $\theta$ is the electron scattering angle, $Q^2 = |\q|^2 -\omega^2$, and 
$\omega$ and $\q$ are the energy and momentum transfer, respectively. 

The structure functions $f$ result from suitable combinations of the components
of the hadron tensor~\cite{Oxford,GP} and depend on the kinematical variables 
$\omega$, $q$, $p'_1$, $p'_2$, on the angles $\gamma_1$, between $\p'_1$ and
$\q$, $\gamma_2$, between $\p'_2$ and $\q$, and $\gamma_{12}$, between $\p'_1$ 
and $\p'_2$. The three structure functions $f'$ are obtained when also the incident
electron is polarized with helicity $h$.

The number of structure functions is reduced in particular kinematics. When the 
angle $\alpha$, between the ($\p'_1$, $\q$) plane and the electron scattering 
plane, is equal to $0$ or $\pi$, the structure functions which are multiplied by 
$\sin\alpha$ or $\sin2\alpha$ in Eq.~(\ref{eq:cssec}) do not contribute to the 
cross section. When $\p'_1$, $\p'_2$, and $\q$ lie all in the same plane, we
recover the case of the (${\vec e},e'N$) reaction and only 5 structure functions
survive: $f_{00}$, $f_{11}$, $f_{01}$, $f_{1-1}$, $f'_{01}$. In a coplanar
kinematics, where both conditions are fulfilled, only the 4 structure functions 
$f_{00}$, $f_{11}$, $f_{01}$, $f_{1-1}$ survive. In the interesting case of the 
super-parallel kinematics, where the two outgoing nucleons are ejected parallel 
and antiparallel to the momentum transfer, only the two structure functions, 
$f_{00}$ and $f_{11}$, survive~\cite{GP}, as in the parallel kinematics of 
($e,e'N$) and in the inclusive ($e,e'$) reaction, and in principle can be 
separated by a Rosenbluth plot.

If one of the two outgoing nucleons has a given polarization, the components of 
the hadron tensor can be written in terms of the spin density matrix, that for 
a nucleon can be expanded in terms of the Pauli matrices $\ss$ as 
\begin{equation}
\rrho = {\frac{1} {2}} \, \Big (1 + {\P}\/\cdot \ss \Big ) , 
\label{eq:rho}
\end{equation}
where ${\P}\/$ is the outgoing nucleon polarization. Thus, the components of the
hadron tensor and therefore also the structure functions can be written as the
sum of two parts, one independent and one dependent on $\P$. As a consequence of
the spin-${\frac{1}{2}}$ nature of the nucleon, the dependence on the spin is at
most linear, and the nine structure functions are written as
\begin{equation} 
f_{\lambda\lambda'} = h^{\mathrm{u}}_{\lambda\lambda'} + {\hat{\s}} \cdot 
\h_{\lambda\lambda'} ,
\label{eq:sfunc}
\end{equation}
where ${\hat {\s}}$ is the unit vector in the spin space. When the outgoing 
nucleon polarization is not detected and the cross section is summed over the 
spin quantum numbers of the outgoing nucleon, the spin independent structure 
functions $h^{\mathrm{u}}_{\lambda\lambda'}$ go over to the 9 structure 
functions $f_{\lambda\lambda'}$ of the unpolarized case~\cite{Oxford,pppol}. New 
structure functions are produced in the polarized case by the components of 
$\h_{\lambda\lambda'}$. 

The quantity $\h_{\lambda\lambda'}$ is usually projected
onto the basis of unit vectors given by $\hat{\L}$\/ (parallel to the momentum 
${\p}\/'$ of the outgoing particle), $\hat{\N}={\q}\/\times{\p}\/'$, and 
$\hat{\S} = \hat{\N}\times\hat{\L}$\/, which define the cm helicity frame of 
the particle. Thus, in the most general situation there are 9 unpolarized 
$h^{\mathrm{u}}_{\lambda\lambda'}$ and 9 polarized 
$h^{\mathrm{i}}_{\lambda\lambda'}$ structure functions for each direction
$i= N, L, S$, i.e. 36 structure functions. 

The coincidence cross section of the (${\vec e},e'{\vec N}N$) reaction can be 
written as
\begin{equation}
\frac{{\mathrm d}^5\sigma} {{\mathrm d}E'_0 {\mathrm d}\Omega'_0  \,
{\mathrm d}\Omega'_1 \/{\mathrm d}\Omega'_2 \/{\mathrm d}E'_1} = 
\sigma_0 \/ {\frac{1} {2}} \, \Big [1 + {\P}\/\cdot \hat {\s} 
+ h(A + {\P}\/'\cdot \hat{\s} )\Big ] ,
\label{eq:cs}
\end{equation}
where $\sigma_0$ is the unpolarized differential cross section, ${\P}\/$ the 
outgoing nucleon polarization, $A$ the electron analyzing power and ${\P}\/'$ 
the polarization transfer coefficient.

The components of ${\P}\/$ and ${\P}\/'$ are given by suitable combinations of 
the polarized structure functions, involving also the matrix elements of the 
lepton tensor, and divided by  $\sigma_0$. The explicit expressions can be found
in~\cite{pppol}. 

In an unrestricted kinematics all the 36 structure functions and all the 
components $P^N$, $P^L$, $P^S$ and $P'^N$, $P'^L$, $P'^S$ are allowed. This 
number is reduced in particular situations. The condition $\alpha = 0$ or $\pi$ 
reduces to 24 the number of structure functions. When the vectors $\q$, $\p'_1$,
$\p'_2$ lie in the same plane, only 18 structure functions survive, as in 
(${\vec e},e'{\vec N}$), and only 12 in coplanar kinematics. In this case  only 
the components $P^N$, $P'^L$ and $P'^S$ survive, as in the coplanar kinematics of 
(${\vec e},e'{\vec N}$). In the super-parallel kinematics, only 5 structure 
functions survive, the 2 unpolarized $h^{\mathrm{u}}_{00}$ and 
$h^{\mathrm{u}}_{11}$ and the three $h^N_{01}, \overline{h}\/'^S_{01}, 
h'^L_{11}$, and thus $P^N$, $P'^L$, and $P'^S$ are each directly proportional to 
only one structure function~\cite{pppol}. 

When final-state interactions (FSI) are neglected and the  plane-wave (PW) 
approximation is used for the outgoing nucleons wave functions, 
$P^N = P^L = P^S = 0$, while $\P'$ does not vanish~\cite{pppol}. 

\section{Formalism for the (${\vec \gamma},{\vec N}N$) reaction}

For the reactions induced by a real photon, only the transverse components of 
the nuclear current contribute and a reduced number of structure functions is 
obtained. The cross section of the (${\vec \gamma},NN$) reaction is written in 
terms of 4 structure functions~\cite{gpnpol}
\begin{equation} 
\frac{{\rm d}^{3}\sigma}{{\rm d}\Omega_{1}{\rm d}\Omega_{2}
{\rm d}E'_{2}} = K_\gamma [  f_{11} 
+ \Pi_{\mathrm c} f'_{11} - \Pi_{\mathrm t} (f_{1-1} 
\cos 2\phi_\gamma -  \bar{f}_{1-1} \sin 2\phi_\gamma) ], \label{eq:csg}
\end{equation}
where $K_\gamma$ is a kinematic factor, $\Pi_{\mathrm c}$ and $\Pi_{\mathrm t}$ 
are the degrees of circular and linear polarization of the photon and 
$\phi_\gamma$ is the angle of the photon polarization vector relative to the 
scattering plane, i.e. the  ($\q, \p'_1$) plane. 

When the photon is not polarized, only $f_{11}$ contributes to the cross section 
$\sigma_0$. Three more structure functions are obtained when the incident photon
is circularly ($f'_{11}$) and linearly ($f_{1-1}$, $\bar{f}_{1-1}$) polarized. 
They can be experimentally separated measuring the circular asymmetry  
\begin{equation}
A_{\mathrm c} = \frac {\sigma(+) - \sigma(-)} {\sigma(+) + \sigma(-)} = 
\frac {f'_{11}} {f_{11}} \, ,
\label{eq:ac}
\end{equation} 
where $\sigma(+)$ and $\sigma(-)$ are the cross sections for a circularly
polarized photon and $\Pi_{\mathrm t}$ = 0, and the two independent linear 
asymmetries,
\begin{equation}
\Sigma_{\frac {\pi} {2}} = \frac {\sigma(0) - \sigma(\frac {\pi} {2})} 
{\sigma(0) + \sigma(\frac {\pi} {2})} = - \frac {f_{1-1}} 
{f_{11}}, \,\,\,\,\, 
\Sigma_{\frac {\pi} {4}} = \frac {\sigma(\frac {\pi} {4}) - 
\sigma(-\frac {\pi} {4})} 
{\sigma(\frac {\pi} {4}) + \sigma(-\frac {\pi} {4})} = \frac {\bar {f}_{1-1}} 
{f_{11}}\, ,
\label{eq:sp2}
\end{equation}
where $\sigma(0)$, $\sigma(\frac {\pi} {2})$ and $\sigma(\frac {\pi} {4})$ are 
the cross sections for a linearly polarized photon in the directions 
$\phi_\gamma = 0, {\frac{\pi} {2}}$, and ${\frac {\pi} {4}}$, 
respectively, and $\Pi_{\mathrm c}$ = 0.

When the polarization of one of the two outgoing nucleons is detected, the 4
structure functions in Eq.~(\ref{eq:csg}) are written, like in 
Eq.~(\ref{eq:sfunc}), in terms of 4 spin independent, 
$h^{\mathrm{u}}_{\lambda\lambda'}$, and 12 spin dependent functions 
$h^i_{\lambda\lambda'}$, 4 for each component $i= N, L, S$. Thus, 
16 structure functions are obtained in the most general situation for the 
(${\vec \gamma},{\vec N}N$) reaction.

The 12 structure functions $h^i_{\lambda\lambda'}$ are directly connected with 
12 polarization observables, 4 for each direction $N$, $L$, and $S$. For an 
unpolarized photon, we have the components of the polarization 
\begin{equation}
P^i = \frac {\sigma(+1) - \sigma(-1)} {\sigma(+1) + \sigma(-1)}
= \frac {h^{i}_{11}} {h_{11}^{\mathrm u}}\, ,
\label{eq:pk}
\end{equation}
where $\sigma(+1)$ ($\sigma(-1)$) are the cross sections with a nucleon
polarized parallel (antiparallel) to the direction $\hat {s}^i$.

If the photon is circularly polarized and $\Pi_{\mathrm t}$ = 0, we have the 
components of the polarization transfer coefficient 
\begin{equation}
P'^i_{\mathrm c} = \frac {\sigma(+,+1) - \sigma(+,-1) - \sigma(-,+1) + 
\sigma(-,-1)} {\sigma(+,+1) + \sigma(+,-1) + \sigma(-,+1) + \sigma(-,-1)}
= \frac {h'^{i}_{11}} {h_{11}^{\mathrm u}}\, ,
\label{eq:pck}
\end{equation}
where the first argument in $\sigma$ refers to the photon polarization 
and the second one to the nucleon polarization in the direction $\hat {s}^i$.

For a linearly polarized photon and $\Pi_{\mathrm c}$ = 0, we have the 
components of the two polarization asymmetries
\begin{equation}
\Sigma^i_{\frac {\pi} {2}} = \frac {\sigma(0,+1) - \sigma(0,-1) - 
\sigma(\frac {\pi} {2},+1) + 
\sigma(\frac {\pi} {2},-1)} {\sigma(0,+1) + \sigma(0,-1) + 
\sigma(\frac {\pi} {2},+1) + \sigma(\frac {\pi} {2},-1)}
= -\frac {h^{i}_{1-1}} {h_{11}^{\mathrm u}}
\label{eq:sp2k}
\end{equation}
and
\begin{equation}
\Sigma^i_{\frac {\pi} {4}} = \frac {\sigma(\frac {\pi} {4},+1) - 
\sigma(\frac {\pi} {4},-1) - \sigma(-\frac {\pi} {4},+1) + 
\sigma(-\frac {\pi} {4},-1)} {\sigma(\frac {\pi} {4},+1) + 
\sigma(\frac {\pi} {4},-1) + 
\sigma(-\frac {\pi} {4},+1) + \sigma(-\frac {\pi} {4},-1)}
= \frac {\bar {h}^{i}_{1-1}} {h_{11}^{\mathrm u}}\, ,
\label{eq:sp4k}
\end{equation}
where the first argument in $\sigma$ refers to the value of the angle
$\phi_\gamma$.

In a coplanar kinematics only 8 structure functions survive: 
$h^{\mathrm u}_{11}$, $h^{\mathrm N}_{11}$, $h^{\mathrm u}_{1-1}$, 
$h^{\mathrm N}_{1-1}$, $h'^{\mathrm L,S}_{11}$, and 
$\bar{h}^{\mathrm L,S}_{1-1}$, and thus only the quantities 
$\sigma_0$, $P^{\mathrm N}$, $\Sigma_{\frac {\pi} {2}}$, 
$\Sigma^{\mathrm N}_{\frac {\pi} {2}}$, $P'^{\mathrm{L,S}}_{\mathrm c}$, and 
$\Sigma^{\mathrm {L,S}}_{\frac {\pi} {4}}$ can be measured. In the PW 
approximation only  $h^{\mathrm u}_{11}$, $h^{\mathrm u}_{1-1}$, 
$\bar{h}^{\mathrm u}_{1-1}$, and $h'^{\mathrm N,L,S}_{11}$ survive, and 
therefore $\sigma_0$, $\Sigma_{\frac {\pi} {2}}$, $\Sigma_{\frac {\pi} {4}}$, 
and $P'^{\mathrm N,L,S}_{\mathrm c}$, while  $A_{\mathrm c}$ = 
$P^{\mathrm N,L,S}$ = $\Sigma^{\mathrm N,L,S}_{\frac {\pi} {2}}$ = 
$\Sigma^{\mathrm N,L,S}_{\frac {\pi} {4}}$ = 0.

All these results can be derived from the general properties of the components
of the hadron tensor~\cite{pppol,gpnpol}.

\section{Results}

Calculations have been performed within the theoretical framework of 
Refs.~\cite{giu98,pn} for exclusive $pp$ and $np$ knockout. In the transition
matrix elements the final-state wave function includes the interaction of each 
one of the two outgoing nucleons with the residual nucleus by means of a 
phenomenological optical potential~\cite{Nad}. The nuclear current is the sum of 
a one-body and of a two-body part, including terms corresponding to the lowest 
order diagrams with one-pion exchange, namely seagull, pion-in-flight, and those 
with intermediate $\Delta$ isobar configurations for $np$ knockout. For $pp$
knockout only the terms corresponding to diagrams with intermediate $\Delta$ 
isobar configurations and without charge exchange contribute to the two-body 
current. 

The two-nucleon overlaps between the ground state of $^{16}$O and the 
lowest-lying discrete final states of  $^{14}$C and $^{14}$N are obtained from 
two  different but similar approaches, based on the Brueckner $G$-matrix for 
$pp$ pairs~\cite{geurts,giu98} and on the coupled cluster method for $np$ 
pairs~\cite{pn}. In both cases the overlaps are expressed in terms of a sum over 
relative and cm wave functions. The coefficients in the sum are obtained in the 
$pp$ case from structure calculations in an extended shell-model space large 
enough to incorporate the corresponding collective features that influence the 
pair removal amplitude. The s.p. propagators used for this description of the 
two-particle propagators also include the effect of both long-range correlations 
and SRC. In the $np$ case the expansion coefficients are determined from a 
configuration mixing calculations of the two-hole states in $^{16}$O, which can 
be coupled to the angular momentum and parity of the requested state and are 
renormalized to account for the spectroscopic factors of the s.p. states. SRC, 
and also tensor correlations for $np$ pairs, are included by the addition of 
state dependent defect functions to the uncorrelated radial partial waves of 
the relative motion. The defect functions were obtained for $pp$ pairs by 
solving the Bethe-Goldstone equation and for $np$ pairs within the framework of 
the coupled cluster method. 

Different types of correlations are contained in the overlap functions, with 
some approximations but consistently. Therefore the spectroscopic factor is 
already included in the calculation. This allows a direct comparison with data, 
without the need to apply a reduction factor, accounting for the correlations 
not included in the model, as it is usually done in the analysis of ($e,e'p$) 
data. 

The results of this model are in reasonable agreement with cross section data
for the $^{16}$O($e,e'pp$)$^{14}$C 
reaction~\cite{Gerco,NIKHEF,Ronald,Rosner,Rosner1}. The comparison has given 
clear evidence for SRC for the transition to ground state of $^{14}$C. 

The experimental separation of the various structure functions in the 
(${\vec e},e'{\vec p}p$) reaction appears extremely difficult. A measurement of 
the components of $\P$ and $\P'$, at least in particular situations, would be 
simpler and less affected by experimental errors, as it is obtained through the 
determination of asymmetries. One can expect that in the polarizations can be 
emphasized effects that are smoothed out in the cross section or that vice 
versa, owing to the ratio, can be smoothed out effects that are important in 
the cross section. Thus, a combined investigation of cross sections and 
polarizations would give complementary information.  

\vfil
\begin{figure}
\epsfysize=12.0cm
\begin{center}
\makebox[16.4cm][c]{\epsfbox{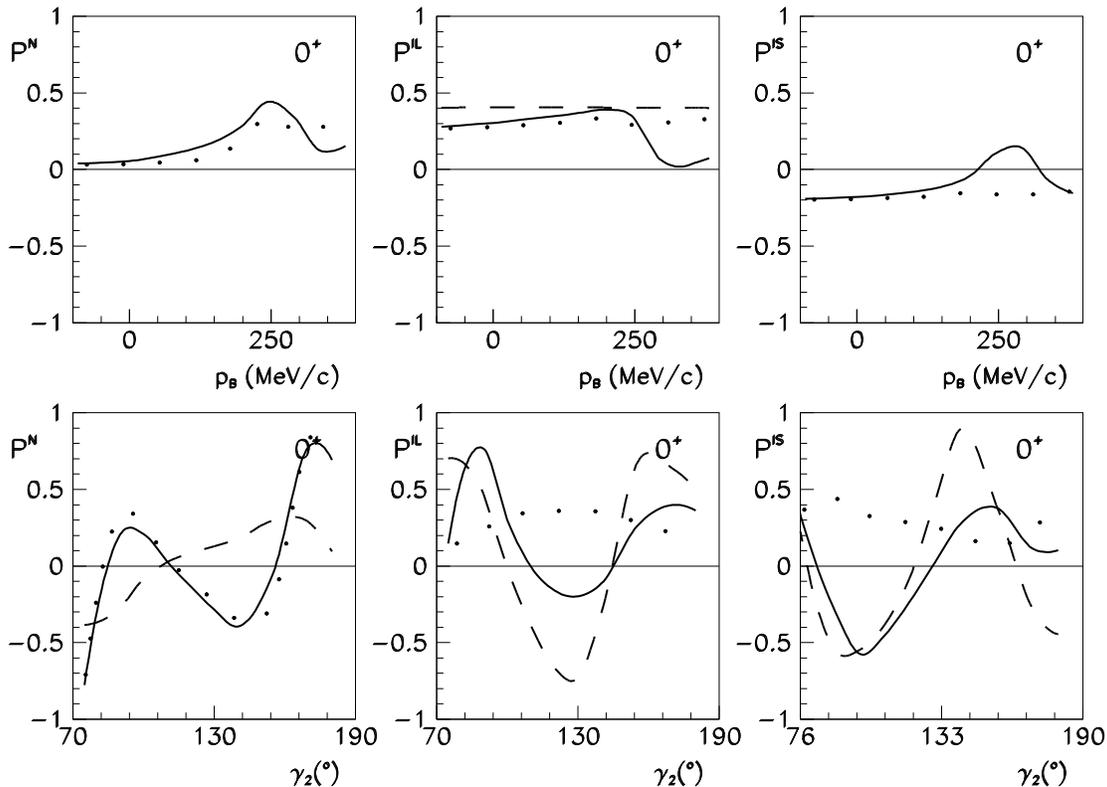}}
\end{center}
    \caption[]{
The components $P^N$, $P'^L$ and $P'^S$ of the 
$^{16}$O(${\vec e},e'{\vec p}p$)$^{14}$C reaction for the transition to the 
$0^+$ ground state of $^{14}$C. Polarization is considered for the proton
characterized by the momentum ${\mbox{\boldmath $p$}}'_1$. The dotted lines 
give the separate contribution of the one-body current, the dashed lines the 
one of the two-body $\Delta$ current and the solid line the total result. The 
defect functions for the Bonn-A $NN$ potential are used. In the upper panels a 
super-parallel kinematics ($\gamma_1=0^{\mathrm{o}}$, 
$\gamma_2=180^{\mathrm{o}}$) is calculated with $E_{0} = 855$ MeV, 
$\omega = 215$ MeV and $q = 316$ MeV/$c$. Different values of  the recoil 
momentum $p_{\mathrm{B}}$ are obtained changing the kinetic energies of the 
outgoing protons. Positive (negative) values of $p_{\mathrm{B}}$ refer to 
situations where ${\mbox{\boldmath $p$}}_{\mathrm{B}}$ is parallel 
(antiparallel) to ${\mbox{\boldmath $q$}}$. In the lower panels a coplanar 
kinematics is considered, with $E_{0} = 584$ MeV, $\omega = 212$ MeV and 
$q = 300$ MeV/$c$, $T'_1 = 137$ MeV and $\gamma_1 = 30^{\mathrm{o}}$, on the 
opposite side of the outgoing electron with respect to the momentum transfer.
\label{fig:lund1}
   }
\vfil\eject
\end{figure}
A numerical example for the components of $\P$ and $\P'$ in the 
$^{16}$O(${\vec e},e'{\vec p}p$)$^{14}$C$_{\mathrm{g.s.}}$ reaction is shown in 
Fig.~\ref{fig:lund1}. Results are compared for two coplanar kinematics 
considered in the cross section measurements at MAMI and NIKHEF. In coplanar 
kinematics only the components $P^N$, $P'^L$ and $P'^S$ survive. The results 
for the super-parallel kinematics in the upper panels are of particular interest, 
because there are plans to measure polarization in this kinematics at 
MAMI~\cite{MAMIpol}. In the super-parallel kinematics, where only one structure 
function contributes to each component, $P^N$, $P'^L$, and $P'^S$ are in general 
smaller than in the calculations displayed in the lower panels, where more 
structure functions contribute. All the polarization components are anyhow 
sizable in both kinematics. 

For the considered transition and in both kinematics the cross sections are 
dominated by the one-body current and thus by SRC~\cite{giu98}. The contribution
of the $\Delta$ current becomes relevant only at large values of 
$p_{\mathrm{B}}$, beyond 200 MeV/$c$. The dominant role of SRC is confirmed in 
Fig.~\ref{fig:lund1} for $P^N$ in both kinematics. Like for the cross section, 
the $\Delta$ current is relevant only at large values of $p_{\mathrm{B}}$. Thus, 
a measurement of $P^N$ would give information on SRC. However, since ${\P}\/$ is
entirely due to FSI, it might be also sensitive to their theoretical treatment.
The role of the $\Delta$ current is more important on  $P'^L$ and $P'^S$,
especially for the kinematics considered in the lower panels.  A measurement  
appears more difficult in this case, since it requires also a polarized incident 
electron. 

\begin{figure}
\epsfysize=12.0cm
\begin{center}
\makebox[16.4cm][c]{\epsfbox{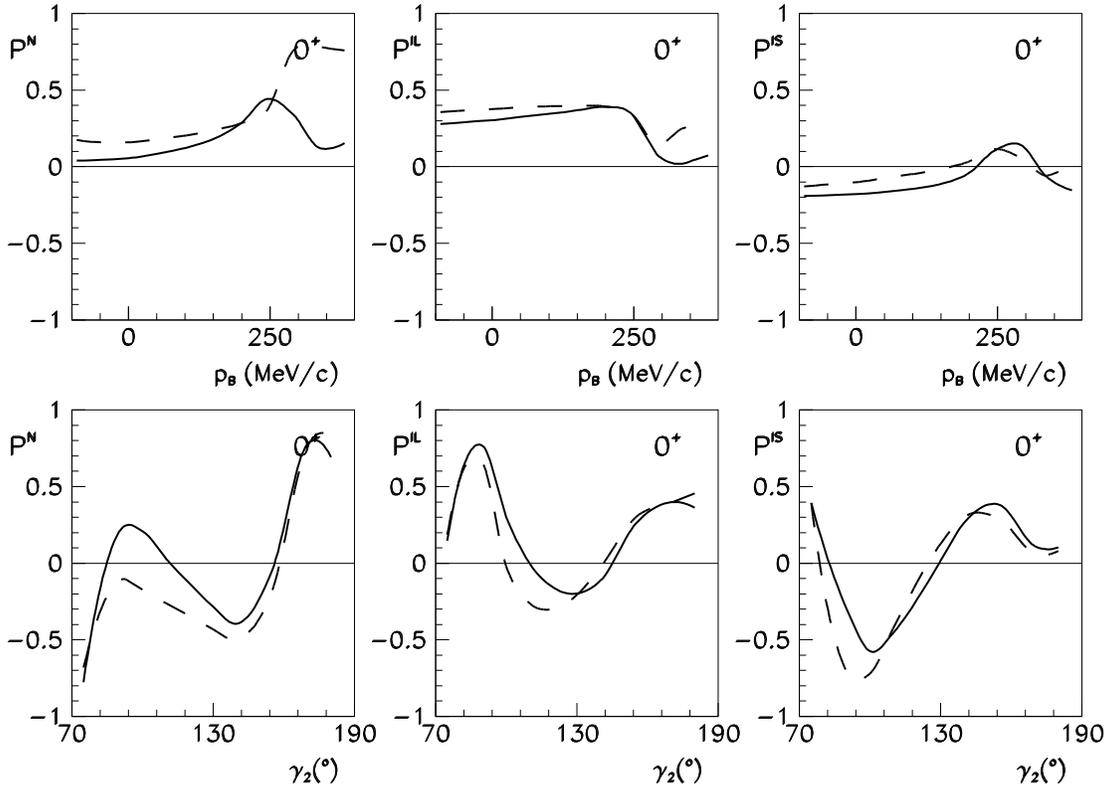}}
\end{center}
    \caption[]{
The components $P^N$, $P'^L$ and $P'^S$ for the same reaction and the same
kinematics in upper and lower panels as in Fig.~\ref{fig:lund1}. The defect 
functions for the Bonn-A and Reid $NN$ potentials are used in the solid and
dashed lines, respectively.
 \label{fig:lund2}
     }
\end{figure}

The results given by the two sets of defect functions from Bonn-A and Reid 
potentials are compared in Fig.~\ref{fig:lund2}. In the considered kinematics 
the two sets give cross sections with the same shape, while the magnitude is 
reduced by a factor of about 2 with Reid~\cite{giu98}. The results for the 
components $P^N$, $P'^L$, and $P'^S$ are qualitatively similar, but there are 
also appreciable differences. This is an indication that cross sections and 
polarization components can have a different sensitivity to the treatment of 
correlations. 
\vfil
\vfil\eject

\begin{figure}
\epsfysize=12.0cm
  \begin{center}
\makebox[16.4cm][c]{\epsfbox{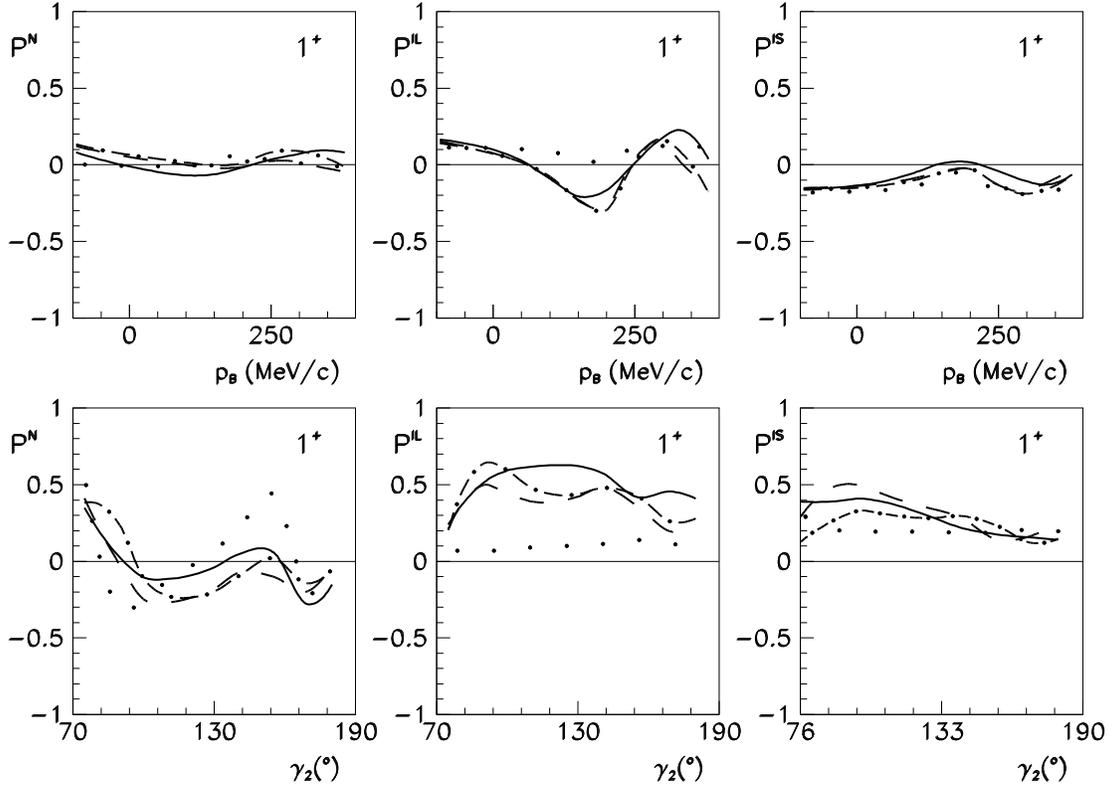}}
  \end{center}
    \caption[]{
The components $P^N$, $P'^L$ and $P'^S$ of the 
$^{16}$O(${\vec e},e'{\vec n}p$)$^{14}$N reaction for the transition to the 
$1^+$ ground state of $^{14}$N. Kinematics in upper and lower panels as in     
Fig.~\ref{fig:lund1}. Defect functions for the Argonne V14 potential are used. 
The separate contributions given by the one-body currents (dotted lines), the 
sum of the one-body and seagull currents (dot-dashed lines), the sum of the 
one-body, seagull and pion-in-flight currents (dashed lines) are displayed. The 
solid lines give the final results, where also the $\Delta$ current is added.
\label{fig:lund3}  }
\end{figure}

In Fig.~\ref{fig:lund3} $P^N$, $P'^L$, and $P'^S$ are displayed for the 
$^{16}$O(${\vec{e}},e'{\vec{n}}p$)$^{14}$N$_{g.s.}$ reaction in the same 
kinematics as in Fig.~\ref{fig:lund1}. The analysis of this reaction would be 
useful for testing the theoretical description of the two-body currents and 
obtaining information on the properties of tensor correlations. Both 
contributions are relevant in this case and the results are influenced by all 
the components of the nuclear current. Correlations are important in particular 
situations, for instance on $P^N$ and $P'^S$ in the super-parallel kinematics, 
where anyhow these quantities are rather small.  
\vfil
\vfil\eject

\begin{figure}
\epsfysize=11.0cm
  \begin{center}
 \makebox[15.4cm][c]{\epsfbox{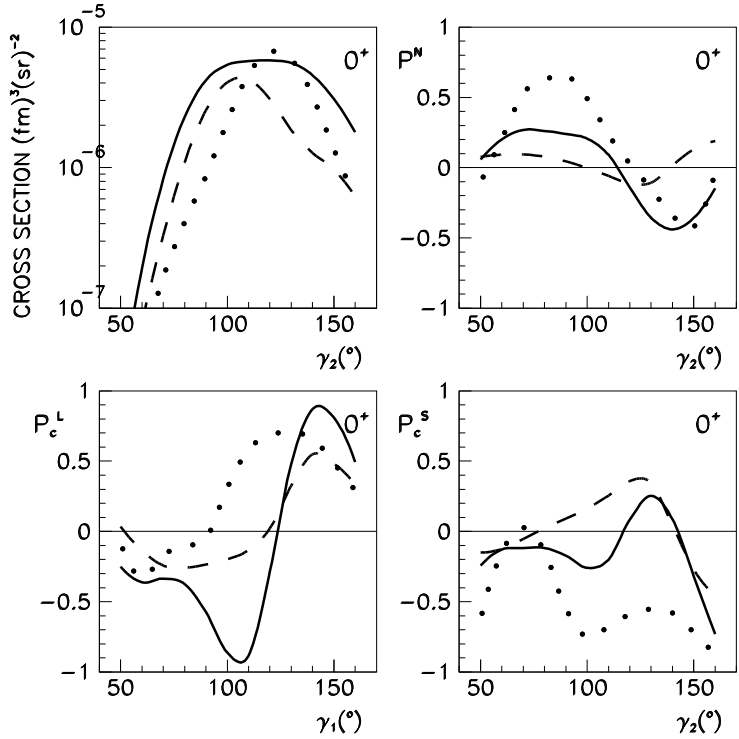}}
  \end{center}
   \caption[]{
The unpolarized differential cross section $\sigma_0$ and the polarization 
observables $P^{\mathrm N}$, $P_{\mathrm{c}}^{\mathrm L}$, and 
$P_{\mathrm{c}}^{\mathrm S}$ of the 
$^{16}$O(${\vec \gamma},{\vec p}p$)$^{14}$C$_{\mathrm{g.s.}}$ reaction in a 
coplanar kinematics with $E_\gamma=120$ MeV, $T'_1 = 45$ MeV and  
$\gamma_1 =45^{\mathrm{o}}$. Defect functions and line convention as in 
Fig.~\ref{fig:lund2}.
 \label{fig:lund4}     }
\end{figure}

\begin{figure}
\epsfysize=11.0cm
  \begin{center}
\makebox[15.4cm][c]{\epsfbox{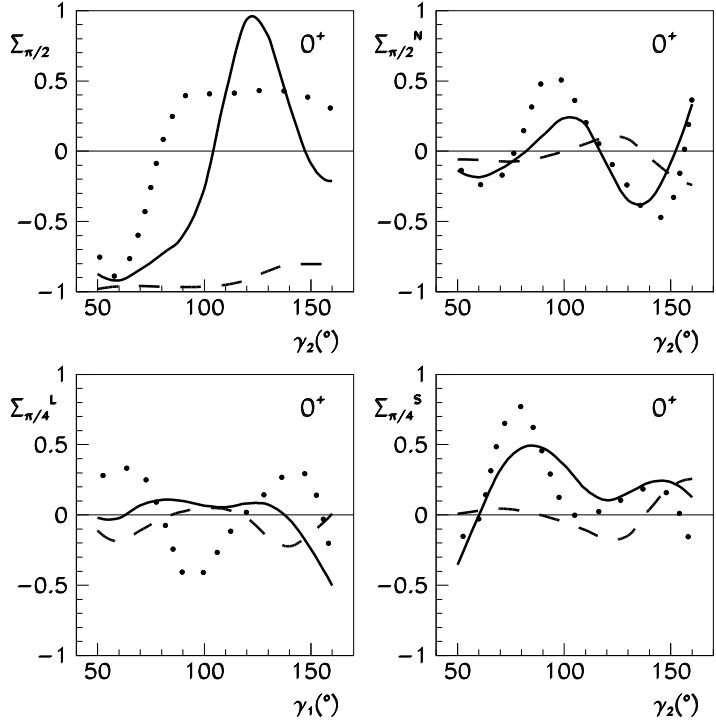}}
\end{center}
    \caption[]{
The polarization observables $\Sigma_{\pi/2}$ 
$\Sigma_{\pi/2}^{\mathrm N}$, $\Sigma_{\pi/4}^{\mathrm L}$, and 
$\Sigma_{\pi/4}^{\mathrm S}$ for the same reaction, in the same conditions and 
kinematics and with the same line convention as in  Fig.~\ref{fig:lund4}.
     \label{fig:lund5} }
\end{figure}

The 16 structure functions obtained in the (${\vec \gamma},{\vec N}N$) reaction
are directly connected with 16 measurable quantities. A complete study requires
out-of-plane kinematics. Some examples can be found in~\cite{gpnpol}. Here a 
simpler coplanar kinematics is considered with an incident photon energy 
$E_\gamma =$ 120 MeV. The calculated unpolarized differential cross section and 
the 7 polarization observables are displayed in  Figs.~\ref{fig:lund4} 
and ~\ref{fig:lund5} for the 
$^{16}$O(${\vec \gamma},{\vec p}p$)$^{14}$C$_{\mathrm{g.s.}}$ reaction. 
The value of the photon energy, far from the peak of the $\Delta$  resonance, 
has been chosen in order to reduce the contribution of the two-body $\Delta$ 
current, that is an interesting but difficult ingredient of the theoretical 
model. Both SRC and two-body currents are important in general. Anyhow either 
process can prevail or can be even dominant for different polarization
observables and in particular kinematics. In Fig.~\ref{fig:lund4} SRC are of
particular relevance on the unpolarized cross section at larger values of 
$\gamma_2$, while  the $\Delta$ gives the major contribution at lower values of 
$\gamma_2$. A similar result is obtained for $P^{\mathrm N}$. The two-body 
$\Delta$ current plays the main role on $P_{\mathrm{c}}^{\mathrm L}$ and 
$P_{\mathrm{c}}^{\mathrm S}$, although the final result is different from the 
two separate contributions.  In Fig.~\ref{fig:lund5} 
$\Sigma_{\pi/2}^{\mathrm N}$ and $\Sigma_{\pi/4}^{\mathrm S}$ are mainly driven 
by SRC, while the $\Delta$ current prevails in $\Sigma_{\pi/4}^{\mathrm L}$. The 
role of SRC is very important also on the photon asymmetry $\Sigma_{\pi/2}$. 
Some of these observables, $P^{\mathrm N}$, $\Sigma_{\pi/2}^{\mathrm N}$, 
$\Sigma_{\pi/4}^{\mathrm L}$, and $\Sigma_{\pi/4}^{\mathrm S}$, vanish in the 
PW approximation and might thus be useful to investigate FSI. This example 
indicates that a combined study of the cross section and polarization obervables 
would provide an interesting tool to unravel and separately investigate the 
different contributions.

\vfil 
\vfil\eject

\vfil
\vfil\eject
This result is confirmed also by the results obtained for the 
$^{16}$O(${\vec \gamma},{\vec n}p$)$^{14}$N reaction~\cite{gpnpol}. In this 
case the role of all the different terms of the two-body current is in general 
more relevant than in $pp$ knockout, but the different sensitivity of the 
various polarization observables to the theoretical ingredients, i.e. to the 
nuclear current, FSI, and to the treatment of $NN$ correlations and nuclear 
structure effects in the two-nucleon overlap function, is confirmed. Moreover, 
different effects can be emphasized in suitable kinematics. 

In conclusion, the study of cross sections and polarizations in electron and 
photon-induced knockout reactions would result in a stringent test of the 
theoretical models and would provide an unique tool to shed light on the genuine 
nature of correlations in nuclei.


\begin{thebibliography}{99}
\bibitem{Oxford} S.~Boffi \etal, Electromagnetic Response of Atomic Nuclei, 
Oxford Studies in Nuclear Physics (Clarendon Press, Oxford, 1996). 
\bibitem{giu98} C.~Giusti \etal, Phys. Rev. \textbf{C 57} (1998) 1691.
\bibitem{pn} C.~Giusti \etal., Phys. Rev. \textbf{C 60} (1999) 054608.
\bibitem{Gerco} C.~J.~G.~Onderwater \etal , Phys. Rev. Lett. \textbf{78} (1997) 
4893.
\bibitem{NIKHEF} C.~J.~G.~Onderwater \etal, Phys. Rev. Lett. \textbf{81} (1998) 
2213.
\bibitem{Ronald} R.~Starink, Phys. Lett. \textbf{B 474} (2000) 33.
\bibitem{Rosner} G. Rosner, Proc. of Conf. on
   ``Perspectives in Hadron Physics'', eds. S. Boffi \etal,
   World Scientific. Singapore (1998) p.185.
\bibitem{Rosner1} G. Rosner, Prog. Part. Nucl. Phys. \textbf{44} (2000) 99.
\bibitem{MAMI} P. Grabmayr \etal, MAMI proposal A1-5/98, Mainz 1998.
\bibitem{Bonn} J. Arends \etal, Z. Phys. \textbf{A 298} (1980) 103.
\bibitem{Tokyo} M. Kanazawa \etal, Phys. Rev. \textbf{C 35} (1987) 1828.
\bibitem{Mainz} J.~C.~McGeorge \etal, Phys. Rev. \textbf{C 51} (1995) 1967;\\ 
P.~Grabmayr \etal, Phys. Lett. \textbf{B 370} (1996) 17;\\
P.~D.~Harty \etal, Phys. Lett. \textbf{B 380} (1996) 247;\\
T. Lamparter \etal, Z. Phys. \textbf{A 355} (1996) 1;\\
I.~J.~D.~MacGregor \etal, Phys. Rev. Lett. \textbf{80} (1998) 245;\\
D.~P.~Watts \etal, Phys. Rev. \textbf{C 62} (2000) 014616.
\bibitem{Lund} L.~Isaksson \etal, Phys. Rev. Lett. \textbf{83} (1999) 3146.
\bibitem{Mainz1} P. Grabmayr \etal, MAMI proposal A2-4/97, Mainz 1997.
\bibitem{Grab} P.~Grabmayr, Proc. of the Fourth Workshop on 
Electromagnetically Induced Two-Nucleon Emission, eds. A.~Lallena and 
P.~Grabmayr, Granada (1999) p. 331.
\bibitem{BGPR} C.~Giusti \etal, Nucl. Phys. \textbf{A 546} (1992) 607; Nucl. 
Phys. \textbf{A 564} (1993) 473.
\bibitem{gnn} C.~Giusti and F.~D.~Pacati, Nucl. Phys. \textbf{A 641} (1998) 297.
\bibitem{Gent2} J.~Ryckebusch \etal, Phys. Rev. \textbf{C 57} (1998) 1318.
\bibitem{LEGSHe} D.~J.~Tedeschi \etal, Phys. Rev. Lett. \textbf{73} (1994) 408.
\bibitem{LEGSO} H.~Baghaei Proc. of the Second Workshop on 
Electromagnetically Induced Two-Nucleon Emission, eds. J.~Ryckebusch and 
M.~Waroquier, Gent (1995) p. 195.
\bibitem{MacGre} I.~J.~D~.MacGregor, Proc. of the Fourth Workshop on 
Electromagnetically Induced Two-Nucleon Emission, eds. A.~Lallena and 
P.~Grabmayr, Granada (1999) p. 339.
\bibitem{pppol} C.~Giusti and F.~D.~Pacati, Phys. Rev. \textbf{C 61}, 
(2000) 054617.
\bibitem{gpnpol} C.~Giusti and F.~D.~Pacati, nucl-th/0102036.
\bibitem{Gent1} J.~Ryckebusch \etal, Phys. Lett. \textbf{B 441} (1998) 1.
\bibitem{GP} C.~Giusti and F.~D.~Pacati, Nucl. Phys. \textbf{A 535} (1991) 573; 
Nucl. Phys. \textbf{A 571} (1994) 694.
\bibitem{geurts} W.~J.~W. Geurts \etal, Phys. Rev. \textbf{C 54} (1996) 1144.
\bibitem{Nad} A.~Nadasen \etal,  Phys. Rev. \textbf{C 23} (1981) 1023.
\bibitem{MAMIpol} M.~Distler, contribution to this workshop.

\end{thebibliography}
\end{document}